\documentclass[twocolumn,aps,prl,showpacs,amsmath,amssymb,10pt]{revtex4-1}
\usepackage{amsmath}
\usepackage{cases}
\usepackage{amssymb}
\usepackage{amsfonts}
\usepackage{txfonts}
\usepackage[dvips]{graphicx}
\usepackage{color}

\begin{document}

\title{Dominant source of disorder in graphene: Charged impurities or ripples?}
\author{Zheyong Fan$^{1,2}$}
\email{Corresponding author: brucenju@gmail.com}
\author{Andreas Uppstu$^{2}$}
\author{Ari Harju$^{2}$}
\affiliation{$^{1}$School of Mathematics and Physics, Bohai University, Jinzhou, 121000, China}
\affiliation{$^{2}$COMP Centre of Excellence, Department of Applied Physics, Aalto University, Helsinki, Finland}
\date{\today}

\begin{abstract}
Experimentally produced graphene sheets exhibit a wide range of
mobility values. Both extrinsic charged impurities and intrinsic
ripples (corrugations) have been suggested to induce long-range
disorder in graphene and could be a candidate for the dominant source
of disorder. Here, using large-scale molecular dynamics and quantum
transport simulations, we find that the hopping disorder and the gauge
and scalar potentials induced by the ripples are
short-ranged, in strong contrast with predictions by continuous
models, and the transport fingerprints of the ripple disorder are very
different from those of charged impurities. We conclude that charged
impurities are the dominant source of disorder in most graphene
samples, whereas scattering by ripples is mainly relevant in the high
carrier density limit of ultraclean graphene samples (with a charged
impurity concentration $\lesssim10$ ppm) at room and higher
temperatures.
\end{abstract}


\maketitle

Although the electronic and transport properties of graphene have been
extensively studied \cite{neto2009,peres2010,mucciolo2010,sarma2011},
a fundamental question of what the dominant disorder limiting the
carrier mobility is has not been definitely answered.  In addition of
theoretical importance, it is vital for graphene-based applications.
Major features of electronic transport in graphene include
\cite{tan2007} a non-universal conductivity at the charge neutrality
point, a density-independent mobility \cite{tan2007,chen2008}, and the
formation of ``electron-hole puddles''
\cite{martin2008,zhang2009}. None of these features can be explained
by short-range disorder that can suppress the conductivity even below
the so-called minimum conductivity $4e^2/\pi h$, via weak and strong
localization \cite{cresti2013,laissardiere2013,fan2014prb}, as
observed in graphene that has been either irradiated \cite{chen2009},
hydrogenated \cite{bostwick2009}, or exposed to ozone
\cite{moser2010}. It is thus most likely that long-range disorder
dominates in graphene.  However, the origin of the long-range disorder
is widely debated.

There are two major candidates for the long-range disorder: extrinsic
charged impurities and intrinsic ripples. Early theoretical works
using the Drude-Boltzmann approach with charged impurities
\cite{hwang2007,adam2007} found good agreement with experimental data
\cite{tan2007}. One the other hand, ripples, observed in several
experiments \cite{meyer2007,ishigami2007}, may also be a major source
of disorder \cite{katsnelson2008,kim2008}. There are experiments
\cite{ponomarenko2009, couto2011} that challenge the viewpoint that
charged impurities are the dominant scatterers, and theoretical
proposals that electron-hole puddles can be induced by the intrinsic
ripples \cite{gazit2009,azar2011,gibertini2012} and that the random
strain fluctuations resulting from the ripples play an important or
even a dominant role in affecting the transport properties of graphene
\cite{couto2014,zwierzycki2014,zhao2015,burgos2015}. The theoretical
works regarding rippled graphene have assumed a continuous model
\cite{vozmediano2010} of long-range diagonal or off-diagonal
disorder. The long-range nature of the disorder induced by the ripples
has not been justified \textit{a priori}, however. In a recent
experimental work \cite{martin2015}, a strong local correlation
between doping and topography in graphene on a metallic substrate, as
predicted theoretically \cite{gazit2009,azar2011,gibertini2012}, was
observed, but the amplitude of the doping was nearly two orders of
magnitude larger than that expected from the theory.

\begin{figure}[htb]
  \includegraphics[width=\columnwidth]{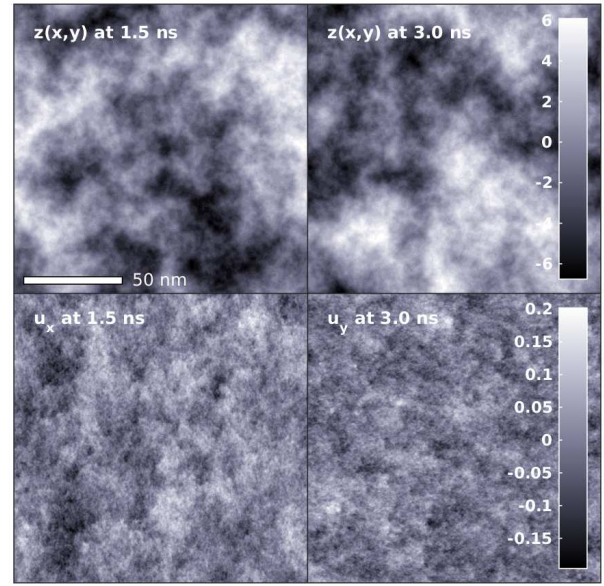}
\caption{Typical MD-relaxed configurations, $z(x, y)$, and in-plane
  displacement fields, $u_x(x,y)$ and $u_y(x,y)$, of a graphene flake
  containing $N=864\,000$ atoms at 300 K. The sample size is about
  $150$ nm $\times 150$ nm and the scale bar is 50 nm. Two
  time points in the MD simulation, 1.5 ns and 3.0 ns, are indicated
  in the figure. The unit of the values in the color bar is \AA.}
  \label{figure:1}
\end{figure}

The aim of this letter is to clarify the dominant disorder limiting the graphene mobility. Omitting the continuous model for the ripples, we perform large-scale molecular dynamics (MD) simulations to obtain realistic graphene ripple configurations, from which we extract the electronic Hamiltonian of large-scale graphene sheets using an accurate tight-binding description that we apply for electronic transport simulations. 
Importantly, we find that the hopping integrals and hence the induced gauge and scalar potentials fluctuate randomly from site to site, forming short-ranged disorder, in strong contrast with the previously used models. The resulting transport properties of this disorder are found to be very different from those of the long-ranged disorder induced by charged impurities. Our results indicate that the charged impurities play a major role of limiting the mobility, apart from a large-carrier-density regime in ultraclean graphene at high temperatures.

The MD simulations are performed using a code implemented on graphics
processing units \cite{fan2015}, with the optimized Tersoff potential
\cite{lindsay2010} tailored for graphene systems.  In the MD
simulations, we start with flat graphene sheets and evolve the systems
by controlling the temperature (adjusting the lattice constant to keep
the in-plane stress zero), ending up with corrugated graphene sheets
after a sufficiently long evolution time of a few nanoseconds.
Typical MD-relaxed configurations of a square-shaped graphene flake
containing $N=864\,000$ atoms at 300 K are shown in the upper part of
Fig. \ref{figure:1}. The height fluctuations $z(x,y)$ are of the order
of a few angstroms, and the characteristic length of the ripple is of
the order of 10 nm, both of which are consistent with experimental
observations \cite{meyer2007,ishigami2007}. The sample-size dependence
of $z(x,y)$ is consistent with that obtained by Monte Carlo
simulations \cite{los2009} and the temperature-dependence of $z(x,y)$
agrees with earlier MD simulations \cite{costamagna2012}. Two
configurations have marginal correlation when they are separated by
about one nanosecond. This means that the ripples are dynamic rather
than static. Apart from out-of-plane deformation, there is also
in-plane deformation, characterized by the potentials $u_x(x, y)$ and
$u_y(x, y)$, defined by the deviation of the corresponding $x$ (the
zigzag direction) and $y$ (the armchair direction) coordinates from
pristine graphene. Typical profiles of $u_x(x, y)$ and $u_y(x, y)$ are
shown in the lower part of Fig. \ref{figure:1}. The in-plane
deformation fields have significantly smaller length and amplitude scales compared to that of the
height fluctuations. It is still crucial for determining the charge carrier disorder.

\begin{figure}
  \centering
  \includegraphics[width=\columnwidth]{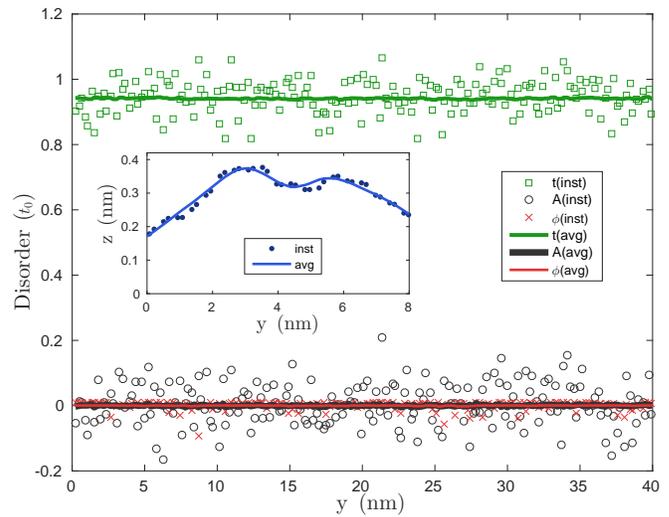}
  \caption{Nearest-neighbour hopping integrals ($t$) and the induced gauge ($A$) and scalar ($\phi$) potentials as a function of the $y$ coordinate along a line with fixed $x$ coordinate in MD-relaxed configurations of a square-shaped graphene flake containing $N=96\,000$ atoms at 300 K. The markers correspond to the results for a fixed-time configuration, while the lines correspond to the results for the average of 500 consecutive configurations separated by a time interval of 1 fs. The inset shows the corresponding instantaneous (markers) and average (line) $z$ coordinate along a shorter line segment.}
  \label{figure:2}
\end{figure}

The fluctuating bond lengths of the rippled graphene lead to modified
hopping integrals and gauge (vector) and scalar potentials for the charge
carriers.  The bond length dependence of the nearest-neighbour hopping
integral $t$ is obtained from a transferable tight-binding model
developed by Porezag \textit{et al.}  \cite{porezag1995}, with the
unperturbed value of $t$ being chosen to be $t_0= -2.7$ eV. The
induced gauge potentials are given by \cite{vozmediano2010} $A_x =
\sqrt{3}(t_2 - t_3)/2$ and $A_y = (t_2+t_3-2t_1)/2$, where the labels
1, 2, and 3 represent the nearest-neighbour atoms corresponding to the
vectors $(-a/\sqrt{3}, 0)$, $(a/2\sqrt{3}, -a/2)$, and $(a/2\sqrt{3},
a/2)$ ($a$ is the lattice constant of graphene). The scalar potential,
which corresponds to on-site potentials and is iduced by local
curvature and rehybridization effects, has been derived to be
\cite{kim2008} $\phi(x,y) \approx -2.3a^2[\nabla^2 z(x,y)]^2$ eV. The
local curvature $\nabla^2 z(x,y)$ can be computed numerically using
second-nearest-neighbour coordinates \cite{azar2011}. Figure
\ref{figure:2} shows the variations of $t$, $A_x$ ($A_y$ behaves
similarly as $A_x$), and $\phi$ along a given line in a sample. It is
clear that both the hopping integral and the induced gauge and scalar
potentials have no long-range correlation. The short-range nature
  of the disorder results from the fact that the deformation fields
  are non-smooth functions of $x$ and $y$ at the atomic scale.

The above results are obtained by considering a single relaxed configuration. One may attempt to average out the short-range fluctuations by a time average of a set of consecutive configurations. According to our MD simulations, the global structure of a fully-relaxed rippled graphene flake does not change over a time scale which is comparable to that of the scattering time of electrons. In view of this, we average over 500 consecutive configurations with an interval of 1 fs between two configurations. This results in smooth deformation fields, but the resulting hopping disorder and random gauge/scalar potentials also disappear, see Fig. \ref{figure:2}. Therefore, one cannot obtain long-range disorder from the ripples even performing a thermal average. The vanishing of the hopping disorder and the resulting gauge potentials under thermal averaging can be understood intuitively, while the vanishing of the scalar potential (which is related to the local curvature)  seems to be counter-intuitive. However, a closer inspection of the thermally-averaged configuration reveals that the in-plane deformation fields do not vanish. The non-vanishing in-plane deformation fields, together with the height fluctuations, result in a structure that is globally corrugated but locally flat. Our atomistic model is in sharp contrast with continuous models, which assume a smooth height function $z(x,y)$ and ignore the in-plane deformation fields, leading to long-ranged gauge and scalar fields.

\begin{figure}
  \centering
  \includegraphics[width=\columnwidth]{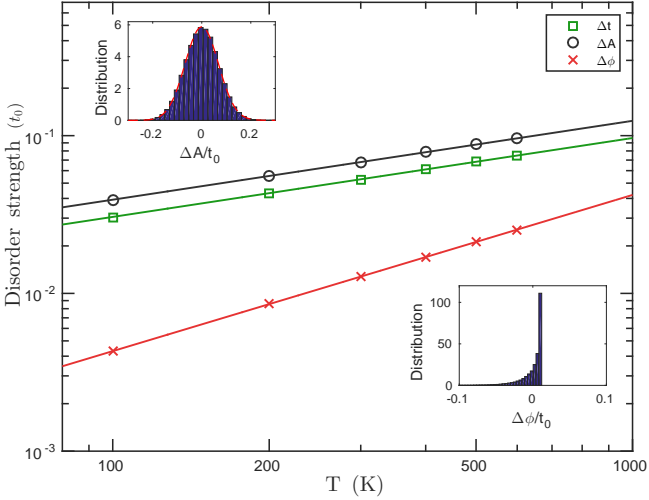}
  \caption{Strengths of the nearest-neighbour hopping disorder $\Delta t$, vector potential $\Delta A$, scalar potential $\Delta \phi$ (defined as the standard deviations) induced by the ripples as a function of temperature. The markers represent the raw data and the lines represent the expressions, $\Delta t/t_0 = 0.053\times \left(T/300 \text{~K} \right)^{1/2}$, $\Delta A/t_0 = 0.068 \times \left(T/300 \text{~K} \right)^{1/2}$, $\Delta \phi/t_0 = 0.013 \times \left(T/300 \text{~K}\right)$,  where $t_0$ is the unperturbed nearest-neighbour hopping. The upper (lower) inset shows the normalized distribution of the vector (scalar) potential in a sample at 300 K. The results here are size-independent.}
  \label{figure:3}
\end{figure}

The gauge potentials follow a Gaussian distribution and the scalar potentials distribute asymmetrically, see insets of Fig.~\ref{figure:3}.
The strengths of the various kinds of disorder can be quantified by their standard deviations: $\Delta t$,
$\Delta A$, and $\Delta \phi$. We find that $\Delta t$ and $\Delta A$
scale as $\sim T^{1/2}$, while $\Delta \phi$ scales as $\sim T$; see
the caption of Fig.~\ref{figure:3} for explicit expressions. The
$T^{1/2}$-scaling of $\Delta t$ and $\Delta A$ can be understood in
terms of the equipartition theorem and the $T$-scaling of $\Delta
\phi$ is related to the second-order coordinate-dependence of the
local curvature $\nabla^2 z(x,y)$. The MD results are not expected to
be accurate at very low temperatures, but they are assumed to be
reasonably accurate above the Bloch-Gr\"{u}neisen temperature
$T_{\text{BG}}$ of the order of 100 K \cite{efetov2010}, where the
phonons follow a quasi-classical distribution. An important
observation from our simulations is that the disorder strength is
largely sample-independent. The disorder strengths shown in
Fig. \ref{figure:3} correspond to a sample with $N=96\,000$ atoms, but
nearly identical results are obtained for other sample sizes as
well. Also, for a given sample, the disorder strengths extracted from
two uncorrelated configurations at different times points in a MD
simulation are also nearly identical. In other words, the disorder
induced by ripples can be well represented by that associated with a
single relaxed sample. This has been confirmed by our transport
calculations, which we now turn to discuss.

After having determined the effect of the ripples to the electronic
structure, we can apply it for large-scale charge transport
simulations.  For these, we use a linear-scaling [$O(N)$] real-space
Kubo-Greenwood method \cite{roche1997} developed for graphics
processing units \cite{fan2014cpc}.  We will compare transport in
rippled graphene with that in graphene with charged impurities. The
Hamiltonian is described in both cases by a tight-binding model $H =
\sum_{i} U_i |i\rangle \langle i| + \sum_{i,j} t_{ij} |i\rangle
\langle j|$, where $U_i$ is the on-site potential at site $i$ and
$t_{ij}$ is the hopping integral between sites $i$ and $j$.  For
rippled graphene the nearest-neighbour hoppings $t_{ij}$ are directly
obtained from MD simulations and the on-site potentials $U_i$ take
values determined by the scalar potential $\phi(x,y)$. Although we
included the scalar potential in our transport calculations, it is
interesting to note that $\Delta \phi \ll 3 \Delta t$ and thus the
off-diagonal disorder $\Delta t$ dominates the carrier scattering in
rippled graphene. For the simulations of charged impurities, we use
pure graphene hoppings $t_{ij} = t_0$ and the local potential $U_i$ is
obtained from the standard model \cite{rycerz2007} of randomly
distributed Gaussian correlated long-range on-site potentials,
i.e. $U_i = \sum_{k=1}^{N_{\text{imp}}} \epsilon_k
\exp\left(-r_{ik}^2/2\xi^2\right)$.  Here $\epsilon_k$ is the strength of
one of the $N_{\text{imp}}$ impurity centres taking a value uniformly
distributed in the interval $[-t_0/4, t_0/4]$, $\xi$ is the effective
range of the charged impurities, and $r_{ik}$ is the distance between
the $i$'th site to the $k$'th impurity centre. Impurity
concentration is defined as $n_{\text{imp}}=N_{\text{imp}}/N$, where
$N$ is the number of atoms in a simulated sample. We use a large
simulation cell of $N=5\times 10^6$ atoms and average the results over
several disorder realizations in all the transport calculations.

\begin{figure}
  \centering
\includegraphics[width=\columnwidth]{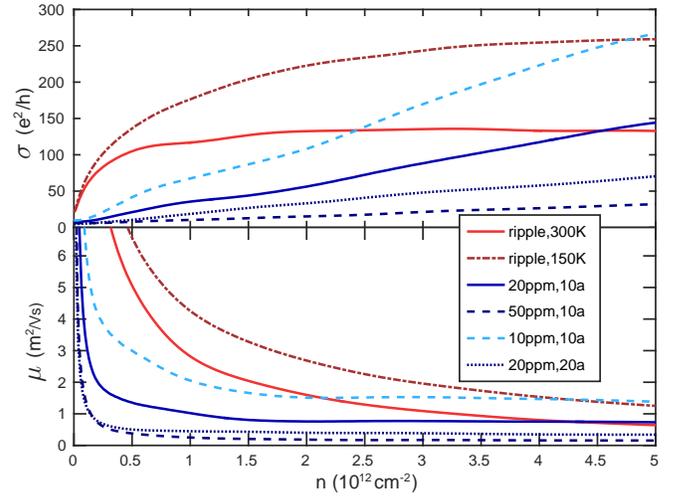}
  \caption{Conductivity $\sigma$ and mobility $\mu$ as a function of the charge carrier concentration $n$. The rippled cases are for 150 K and 300 K and charged impurity cases are labelled by the defect density ($n_{\text{imp}} =10-50$ ppm) and range ($\xi=10-20$a).}
  \label{figure:4}
\end{figure}

Figure \ref{figure:4} shows the conductivity $\sigma$ calculated as a function of the carrier concentration $n$ both in graphene with ripples and in graphene with charged impurities. 
The rippled sample at 300 K shows a clear saturation of the conductivity to a constant value at large $n$. 
This is  in strong contrast with results for charged defects showing unequivocal linear increase of the conductivity as a function of $n$, for all the defect densities and potential ranges used. 
Using the Einstein relation of diffusion, the corresponding mean free path (MFP) $l$ can be computed as $l = 2\sigma/e^2\rho(E)v_F$, where $v_F$ is the Fermi velocity, 
$\rho(E)=2E/\hbar^2v_F^2$ is the electron density of states (spin degeneracy included). Converting density of states to carrier density $n(E)=\int_0^E\rho(E)dE=E^2/\hbar^2v_F^2$, 
we get $l = \sigma n^{-1/2}/(e^2/\hbar)$.
Therefore, the MFP resulting from ripples $l_{\text{ri}}$ is
proportional to $n^{-1/2}$ and for charged defects $l_{\text{cd}} \sim
n^{1/2}$.  The scaling of $l_{\text{ri}}$ happens to be the same as
that for electron-phonon scattering in the high-temperature regime
obtained in Ref.~\cite{li2013}. This suggests that the
temperature-induced random hopping disorder captures essential effects
of the electron-phonon scattering. Indeed, based on the $T^{1/2}$
scaling of $\Delta t$ and Fermi's golden rule, we have the
inverse-temperature scalings, $l_{\text{ri}} \sim T^{-1}$. Results for
rippled graphene at a lower temperature of 150 K are also shown in
Fig. \ref{figure:4} for comparison \cite{note:ballistic}. This trend
is in good agreement with the linear increase of resistivity with
increasing temperature from 50 K to 240 K in suspended graphene
\cite{bolotin2008}.  Actually, this linear scaling is found to be
valid down to $T=0.2T_{\text{BG}}$ \cite{efetov2010}, where the
Bloch-Gr\"{u}neisen temperature $T_{\text{BG}}$ is about 100 K
\cite{efetov2010} in the low-carrier-density regime.
Based on our numerical results, we can also obtain a practical formula expressing the electron-phonon scattering MFP mediated by the ripples as a function of temperature and carrier density: $l_{\text{ri}}(T, n) \approx 20/ (T/300 \text{ K})^{-1} n^{-1/2}$. At $T=300$ K, $l_{\text{ri}}$ is about 200 nm at a moderate carrier density of $n=10^{12}$ cm$^{-2}$, but can be 2 $\mu$m at a small carrier density of $n=10^{10}$ cm$^{-2}$. This formula represents an upper limit of the electron MFP achievable in suspended graphene.

The conductivity can be converted to mobility $\mu$ using the Drude model $\mu=\sigma/en$.  
Away from the charge neutrality point, the charged defects lead to a $\mu_{\text{cd}}$ that is largely independent of $n$, whereas $\mu_{\text{ri}}$ for the rippled case shows a clear drop as a function of the carrier density. 
Quantitatively, the ripple-induced mobility is $\mu_{\text{ri}} \approx 20 (e/\hbar) (T/300 \text{ K})^{-1} n^{-1}$. At $T=300$ K, $\mu_{\text{ri}}$ is about 3 $\text{m}^2/\text{Vs}$ at a moderate carrier density of $n=10^{12}$ cm$^{-2}$, and can be 300 $\text{m}^2/\text{Vs}$ at a small carrier density of $n=10^{10}$ cm$^{-2}$.
We can thus conclude that the thermal rippling of graphene does not result in similar transport properties as caused by the long-range scatterers.
The results for long-range charged defects are in good agreement with experimental findings \cite{tan2007,chen2008} for supported graphene.

\begin{figure}
  \centering
\includegraphics[width=.99\columnwidth]{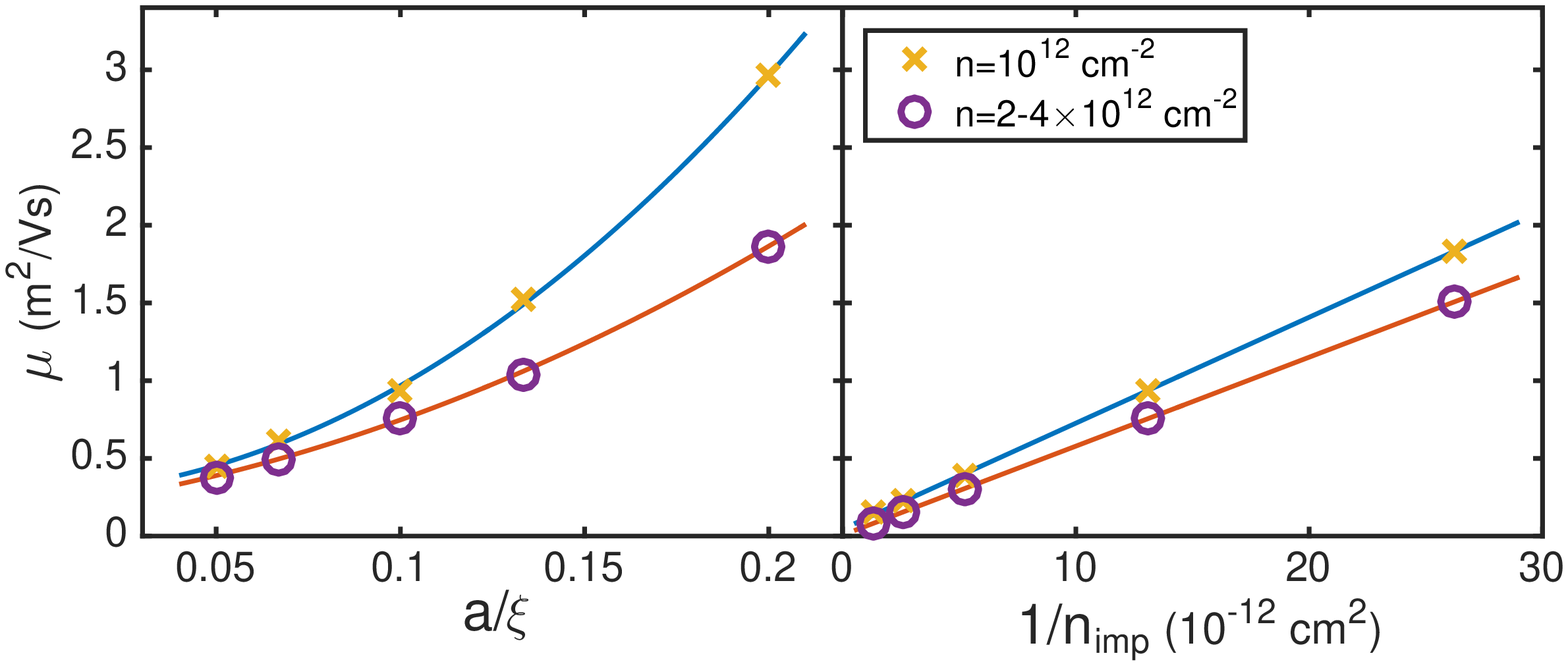}
  \caption{Scaling of the mobility of graphene with charged defects as
    a function of the inverse impurity range ($\xi$) or density
    ($n_{\text{imp}}$). Lines are a guide to the eye.}
  \label{figure:5}
\end{figure}

The scaling of the mobility of graphene with charged impurities as a
function of impurity density $n_{\text{imp}}$ or range $\xi$ is shown
in Fig. \ref{figure:5}. It is shown for a fixed charge carrier density
of $10^{12}$cm$^{-2}$ and for an average of a region ranging from
double to four times the fixed value. The fixed value corresponds to a
modest and the average to a large charge carrier regime.  The
dependence on the impurity density is inversely linear, whereas the
mobility shows quantitatively a $\sim 1/\xi^2$ dependence on the
impurity range, which can be understood by the fact that the size of
the disordered region scales as $\sim\xi^2$. These scaling behaviours
can be used to compare our findings with different experimental
results.

For cases where several scattering sources are present, the mobilities caused by each scatterer type $\mu_i$ can be approximately combined using Matthiessen's rule as $1/\mu= \sum_i 1/\mu_i$.
The mobility in rippled graphene at 300 K is about 3 $\text{m}^2/\text{Vs}$ when $n=10^{12}\text{cm}^{-2}$, as noted above, which is comparable to that in graphene with charged impurities at a concentration of $n_{\text{imp}}=10$ ppm. 
According to the inverse dependence of $\mu$ on $n_{\text{imp}}$, one can infer that when $n_{\text{imp}}$ is reduced to about $10^{10}\text{cm}^{-2}$, the thermal ripples would become the major source of disorder around and above room temperature. This, however, excludes ripples as the major source of disorder in graphene samples on substrate. 

In suspended graphene, the dominant source of disorder may also not be ripples if there is a relatively high concentration of adsorbed charged impurities. A rapid increase of $\mu$ has been observed by 
Newaz \textit{et al}. \cite{newaz2012} in graphene samples suspended in liquids at room temperature when the dielectric constant $\kappa$ of the surrounding media is increased from 1.9 to 4.3, which can be caused by a decrease of the charged impurity range $\xi$. A further increase of $\kappa$, however, brings $\mu$ back to lower values, which was attributed to additional scattering caused by ions present in the polar liquid ($\kappa>5$) \cite{newaz2012}. The same mechanism may account for the insensitivity of the mobility to the dielectric environment of supported graphene as observed in earlier works \cite{ponomarenko2009, couto2011}, which has challenged the dominant role of charged impurities in supported graphene. 

To conclude, we have shown that although the thermal corrugation of graphene is a long-wavelength phenomenon, the dominant disorder caused by ripples affecting the charge carriers is rather counterintuitively very similar to that caused by short-range scatterers. Our electronic transport simulations demonstrate clear differences in the conductivity and mobility of graphene with ripples and with long-range charged defects. Based on our transport data, the properties of graphene with long-range charged defects correlate strongly with experimental findings whereas scattering by ripples is mainly relevant for ultraclean graphene samples at elevated temperatures.

\begin{acknowledgments}
This work was supported by the Academy of Finland through its Centres of Excellence Programme (2015-2017) under project number 284621.
We acknowledge the computational resources provided by Aalto Science-IT project and Finland's IT Center for Science (CSC). Z. Fan was supported by NSFC
(Project No. 11404033).
\end{acknowledgments}

\end{document}